\documentclass{aastex}
\usepackage{spr-astr-addons}
\usepackage{url}

\RequirePackage{color}

\begin{document}

\title{Multifractality due to long-range correlation in the L-band ionospheric scintillation $S_4$ index time series }
\slugcomment{Not to appear in Nonlearned J., 45.}
%% Running heads
\shorttitle{Short article title}
\shortauthors{Autors et al.}

\author{H.J.Tanna\altaffilmark{*}} \and \author{K.N.Pathak}
\affil{Department of Applied Physics, S. V. National Institute of  Technology, Ichchhanath, Surat- 395007, Gujarat, India.}
\email{hemalitanna8@gmail.com}
\altaffiltext{*}{E-mail : hemalitanna8@gmail.com; t.hemali@ashd.svnit.ac.in\  Tel.:+91 9904773706}

\begin{abstract}
The earth's ionosphere is well recognized as a dynamical system and non-linearly coupled with the magnetosphere above and natural atmosphere below. The shape and time variability of the ionosphere indeed shows chaos, pattern formation, random behaviour and self-organization.The present paper studies the propriety of Multifractal Detrended Fluctuation Analysis (MF-DFA) technique for the ionospheric scintillation index time series. MF-DFA is used to identify the scaling behaviour of the ionospheric scintillation time-series data of two different nature.The obtained results show the robustness and the relevancy of the MF-DFA technique for the ionospheric scintillation index time series. The comparison of the MF-DFA results of original data to those of shuffled and surrogate series shows that the multifractal nature of considered time-series is almost due to long-range correlations. Subsequently, the Hurst exponents derived from two parallel methods namely Rescaled range analysis (R/S) and Auto Correlation Function (ACF) are also suggesting the presence of long range correlation. The presented results in this work may be of assistance for future modeling and simulation studies.
\end{abstract}


%\section*{}
%\label{sec:intro}

\section{Introduction}

The history of ionospheric research, starting with the pioneering experiments by Appleton and Barnett (1925) and Breit and Tuve (1926), is long and rich in physics and chemistry (Rishbeth et al. 1996). The Earth's ionosphere is composed of energy and neutral atoms/molecules for ionization process and is well recognized as a complex non-linear system. Being governed by classical equations, the real ionosphere is a dynamical system of non-linearly coupled fields to the magnetosphere above and neutral atmosphere below and strongly affected by neutral winds, atmospheric tides, motion due to electric and magnetic fields etc. Indeed, the space and time variability of the real ionosphere shows chaos, pattern formation, random behaviour and self-organization (Materassi et al. 2003). One of the most dramatic manifestations of the "irregularity" of the ionosphere is the fluctuation of radio signals crossing it. The rapid fluctuation of amplitude and phase of the radio signal passing through the ionospheric region, embedded with plasma density irregularities refers to as the ionospheric scintillation.

The irregularities that can cause scintillation of radio waves are being studied using data from digisonde, scanning photometers/imager (optical techniques) and VHF monitors among others (Oladipo and Sch$\ddot{u}$ler 2013). Ionospheric scintillations mainly occur near the magnetic equator region essentially at night, shortly after local sunset and are associated with the presence of spread-F irregularities in the path of radio waves. However, there is a recent agreement that day-time ionospheric scintillations are sometimes associated with E region irregularities (Zou and Wang 2009; Zou 2011). The existence of ionospheric scintillation degrades the performance of satellite based communication system, resulting in signal fading below the fad margin of the receiver, and leading to the signal loss and cycle slips (Singh et al. 2006; Seo et al. 2011). This renders a study of ionospheric scintillation interesting and challenging for both space scientists and radio communication systems engineers.

Because of its temporal and spatial availability, GPS has been considered as an ideal system for studying the L-band ionospheric scintillation (Oladipo and Sch{\"u}ler 2013). In the case of GPS, the amplitude scintillation is basically recorded as an index, in which the most commonly used is $S_4$index (Muella et al. 2008). By considering the scintillation ($S_4$ index) level threshold of 0.2, the morphological characteristics of ionospheric scintillation at low latitudes in different longitude regions are extensively studied by many researchers (Paula et al. 2003; Gwal et al. 2006; Shang et al. 2008; Li, et al. 2008; Muella et al. 2008). The statistical properties of the $S_4$ index data on different time scales can be a unique tool to identify the parameters of the standard model of ionospheric scintillation. Moreover, there is evidence that, scintillation theory relates the observed signal statistics to the statistics of ionospheric electron density irregularities. It has been found for instance, that phase and amplitude spectra can be used to deduce the scintillation spectrum of electron density fluctuations (Yeh and Liu 1982). However, Costa and Kelley (1978) and Bhattacharya (1990) have reported that power spectrum does not describe unambiguously the turbulent nature of the ionosphere. For the characterization of ionospheric turbulence, some alternative nonlinear analysis methods have been suggested by Wernik et al. (2003) which are wavelet transform, higher order moments, fractal and multifractal analysis, etc.. An attempt to obtain statistical features of ionospheric irregularities and its generation mechanism at low latitude station Varanasi, the spectral analysis of VHF scintillation has been made by Singh et al. (2006); they have also reported that for studying the observation of the turbulent nature of the ionosphere, an accurate nonlinear theory must be developed. Recently, Unnikrishnan and Ravindran (2010) and Unnikrishnan (2010) have studied the chaotic behavior of equatorial/low latitude ionosphere through the deterministic chaotic analysis of GPS TEC time series.

Many of the time series from complex systems exhibit self similarity, which is the signature of a fractal nature in the
system (Feder 1988). There are several approaches to analyze the self similarity/fractality in the time series, such as Autocorrelation Function (ACF), Spectral analysis, Hurst's Rescaled-Range analysis (R/S) and Fluctuation analysis, that provide type of self affinity for stationary time series (Kantelhardt 2008). The definition of the fractal is associated to the ability of a system to present similar features with a change of scale, reflected in the response signal which has similar and reproducible statistical features. Fractals can be classified into two categories: monofractals and multifractals. Monofractals are those, whose scaling properties are the same in different regions of the system meaning that, one single scaling exponent, called global Hurst exponent H is sufficient to describe its scaling properties. However, in natural fractals, the self-similarity does not hold in a single scaling exponent but it holds statistically over a finite range of scales i.e. multifractal (Hu et al., 2001; Chen et al., 2002; Kantelhardt et al., 2002). Thus, a fundamental characteristic of multifractal structure is that the scaling properties may be different in different segments of the system requiring more than one scaling exponent to be completely described.

The simplest type of multifractal analysis is based upon the standard partition function multifractal formalism, which has been developed for the multifractal characterization of normalized, stationary measurements (Feder 1988; Barabasi and Vicsek, 1991; Peitgen et al., 1992; Bacry et al., 2001). However, this standard formalism does not give correct results for non-stationary time series that are affected by trends or that cannot be normalized. Therefore, an improved multifractal formalism was developed known as wavelet transform modulus maxima (WTMM) method (Muzy et al., 1991), which is based on the wavelet analysis and involves tracing the maxima lines in the continuous wavelet transform over all scales. At the Brazilian sector, Bolzan et al. (2009) and Bolzan et al. (2013) have used this WTMM method to get the multifractal spectra of TEC time series. Peng et al. (1995) have introduced the Detrended Fluctuation Analysis (DFA) Method, to study the properties of DNA sequences. Since its introduction, this method has been widely used in diverse fields of solar activities to the earth science (e.g. Geology, DNA sequences, neuron spiking, heart rate dynamics, economic time series and also weather related and earthquake signals) (Arneodo et al., 1995; Molchanov and Hayakawa, 1995; Hayakawa et al., 1999; Kantelhardt et al., 2001; Telesca et al., 2001; Varotsos et al., 2002; Varotsos et al., 2003; Vyushin et al., 2004). By generalizing this standard DFA method, Kantelhardt et al. (2002) have introduced the Multifractal Detrended Fluctuation Analysis (MF-DFA), which allows the global detection of multifractal behavior. By applying this MF-DFA method to the sunspot number time series, Movahed et al. (2006) and Hu et al. (2009) have found that the presence of multifractality/complexity in the sunspot number fluctuations is almost due to long range correlation. Several researchers have also studied the multifractal characteristics of earthquake related signals by the MF-DFA method (Biswas et al. 2012; Ghosh et al. 2012; Masci 2013).

In this work we characterize the complex behavior of ionospheric dynamics through the computation of the fluctuation
parameters$-$scaling exponents which quantifies the correlation exponents and multifractality of the data sets. In order to study the behavior of ionospheric dynamics, the day time (scintillation free period) and night time (highly scintillated) $S_4$ index time series have been used in the present work. The $S_4$ index time series is obtained from the recorded data of the GSV4004B GISTM (GPS Ionospheric Scintillation TEC Monitor) receiver at an Indian low latitude station Surat (21.16$^{0}$N, 72.78$^{0}$E; Geomagnetic: 12.90$^{0}$N, 147.35$^{0}$E). We use fractal analysis approaches, such as  R/S,  ACF and MF-DFA to analyze the data set. We investigate persistency in the $S_4$ index time series by calculating the Hurst exponent through the R/S and ACF methods. To identify the multifractality in the ionospheric dynamics, the MF-DFA have been applied to the day time  and the night time  data sets. An algorithm of MF-DFA, basically calculates the fluctuation function Fq(s), generalized Hurst exponent h(q), multifractal scaling exponent $\displaystyle \tau$(q), and multifractal spectrum f($\alpha$) (Kantelhardt et al. 2002). By comparing MF-DFA results of the original scintillation index time series to those for shuffled and surrogate series, we show that the multifractal nature of ionospheric fluctuations is mainly due to long-range correlation.

The paper is organized as follows. In the following section, we describe the data and methodology. The required steps for producing R/S, ACF and MF-DFA methods are introduced in Section 3. Our results are presented and discussed in Section 4 and, finally, the concluding remarks of this work are summarized in Section 5.

\section{Data and methodology}

The GSV4004B receiver tracks up to 11 GPS satellites at the L1 (1575.42MHz) and L2 (1227.60MHz) frequency simultaneously. It records phase and amplitude at a 50-Hz rate for each satellite tracked on L1 and calculates amplitude and phase scintillation parameters in real time. The amplitude scintillation index, $S_{4}$, is computed over 60-s intervals as the standard deviation of the detrended signal intensity (SI) normalized by its mean value and is referred to as the Total $S_{4}$. The GSV4004B receiver computes the Total $S_{4}$ ($S_{4T}$) and the correction to the Total $S_{4}$ ($S_{4N0}$) due to ambient noise over 60s intervals in real time. The corrected $S_{4}$ with the effects of ambient noise removed is then computed as follows (Zou et al. 2009):

\begin{equation}
S_{4} = \sqrt{S^{2}_{4T}-S^{2}_{4N0}}\nonumber
\end{equation}

To confirm consistent statistics, following two criteria (Tanna et al. 2013) have been used in present analysis:

1. In order to minimize the effects of the multipath on the observations, measurements with a satellite's elevation angle greater than $30^0$ are taken into account and

2. The analysis is limited to measurements made from satellites that are locked on for greater than 4 min (240 seconds) to allow the GSV4004B receiver's detrending  filter to stabilize, as it has to be reinitialized whenever the lock to the carrier phase is lost.
\section{Analysis methods}
In this section we present three standard methods: R/S, ACF and MF-DFA, to investigate the fractal properties of inhomogeneous ionosphere.
\subsection{Rescaled Range (R/S) Analysis}
The rescaled range (R/S) analysis is a simple but a strong nonparametric method for fast fractal analysis (Das et al. 2009). This is performed on the discrete time series data set X(i) of dimension N by calculating three factors: first, the mean  $\langle$X$\rangle$ , second the R, which is the range of cumulative difference of X (t, s) at discrete time t over time span s, and third, the standard deviation S(i), estimated from the observed values of X(t, s). Where,
\begin{equation}
\langle{X}\rangle=\frac{1}{N}\sum^{N}_{i=1}{X(i)}
\end{equation}
and
\begin{equation}
S(i)= \{{\frac{1}{s}\sum^{s}_{i=1}{[X(i)-\langle{X}\rangle}]^{2}}\}^{\frac{1}{2}}, s=1...N
\end{equation}
The Range of cumulative difference of the data is given by
\begin{equation}
R(i)= Max[Y(i)]-Min[Y(i)]
\end{equation}
Where, Y (i) is defined as follows
\begin{equation}
Y(i)= \sum^{i}_{k=1}{[X(k)-\langle{X}\rangle]}
\end{equation}

Hurst found that the ratio (R/S) is very well described for the large number of natural phenomena by the following experimental relation (Hurst 1951):
\begin{equation}
\langle{R/S}\rangle = s^{H}
\end{equation}
where, s and H are time span and Hurst exponent, respectively. The slope of the plot of  R/S  versus the time span s on log-log plot gives rise to H. The value of H indicates whether a time series is random or successive increments in time series are not independent. The fractal dimension, D is determined from H as D = 2-H (Das et al. 2009). The correlation between two successive steps or increments was represented by:   $\rho$ = $2^{2H-1}$-1 (Feder 1988). For H equals to 0.5 ( $\rho $= 0) the time series represents a random walk or uncorrelated and each observation is fully independent of all prior observations. H lying between 0.5 and 1 ( $\rho$ positive) implies persistent time series characterized by long range effects. The successive increments are positively correlated with the preceding observations. For 0 $<$ H $<$ 0.5, ($\rho$  negative) the exponents indicate anti-persistent and each data points are more likely to have a negative correlation with preceding values (Das et al. 2009).
\subsection{Auto-Correlation Function (ACF)}
For the time series of dimension N, Xi, i = 1, 2, 3, ���., N, the ACF can be written as (Barbieri and Vivoli 2005)
\begin{equation}
C(s)= \frac{\frac{1}{N-s}\sum^{n-s}_{j=1}{(X_{j+s}-\langle{X}\rangle)(X_{j}-\langle{X}\rangle)}}{\frac{1}{N}\sum^{N}_{j=1}{(X_{j}-\langle{X}\rangle)}^{2}}
\end{equation}
where s and  X  are the time span and mean of the time series respectively. If there is long-range time dependence in the natural phenomena, then the algebraic decay of the ACF can be described by the following power law relation C(s)= $s^{-\alpha}$, with an ACF exponent 0 $<$$\alpha$$<$ 1 for large s  (Rangarajan and Ding 2000).
The long-term persistent data can be defined by the relation (Rangarajan and Ding 2000; Nurujjaman and Iyengar 2007):
\begin{equation}
H= 1- \frac{\alpha}{2}.
\end{equation}
The correlation exponent $\alpha$ and Hurst exponent H can thus be obtained by calculating the slope of double logarithmic plot of C(s) versus time span s.

\subsection{Multifractal Detrended Fluctuation Analysis (MF-DFA)}

A Multifractal Detrended Fluctuation Analysis (MF-DFA), is based on the identification of the scaling of the $q^{th}$ order moments depending upon signal length and is a generalization of the standard DFA using only the second moment q=2.  An algorithm of MF-DFA prepared in LABVIEW, basically consists of five steps (Kantelhardt et al. 2002).

Let X (i) be the time series of amplitude scintillation index $S_4$ for i = 1,......, N, N being the length. Here, the time series X(i) is assumed to be the increments of a random walk process around the average thus, in the first step the integrated profile Y (i) is obtained by the Eq. (4).

In the second step, the integrated time series obtained from Eq. (4) is divided to Ns non-overlapping bins, where Ns = int (N/s) and s is the time span. Since N is not multiple of s, a short part of the series is left at the end. So in order to include this part of the series we have repeated the entire process starting from the opposite end, thus leaving a short part at the beginning. Thus, we have obtained 2Ns bins.

In the third step, for each one of the Ns bins, we perform the polynomial fit of the data and then estimate the variance for each bin v, v = 1,..., Ns and v = Ns +1, ...,2Ns from the following Eq. (8) and (9) respectively.
\begin{equation}
F^{2}(s,v)=\frac{1}{s}\sum^{s}_{i=1}\{{Y(i)[(v-1)s+i]-Y_{v}(i)}\}^{2}
\end{equation}
\begin{equation}
F^{2}(s,v)=\frac{1}{s}\sum^{s}_{i=1}\{{Y(i)[N-(v-N_{s})s+i]-Y_{v}(i)}\}^{2}
\end{equation}
Where, Y(i) is the polynomial fitted values in the bin v.
In order to analyze the influence of fluctuations of different magnitudes and on different time spans, the $q^{th}$ order fluctuation function given by
\begin{equation}
F_{q}(s)=\{\frac{1}{2N_{s}}\sum^{2N_{s}}_{v=1}[F^{2}(s,v)]^{\frac{q}{2}}\}^{\frac{1}{q}},
\end{equation}
is obtained for a real valued parameter q $\neq$ 0 in the fourth step. For q=0, a logarithmic averaging procedure has to be applied because of the diverging exponent
\begin{equation}
F_{o}(s)=exp\{\frac{1}{4N_{s}}\sum^{2N_{s}}_{v=1}ln[F^{2}(s,v)]\} \sim s^{h(0)}.
\end{equation}
For q = 2, the standard DFA procedure is retrieved by construction Fq (s), is only defined for s $\geq$ m+2. As the main purpose is to ascertain the scaling behaviour, the generalized fluctuation functions should be estimated for different values of the time spans s and for different values of the order q.  Generally, if the time series X(i) has long-range power-law correlations, then Fq(s) increases for large values of scale s as a power-low i.e.,
\begin{equation}
F_{q}(s)\sim s^{h(q)}.
\end{equation}
Hence in the last step, the scaling exponent h(q), usually known as generalized Hurst exponents, are estimated by analyzing the double logarithmic plot of Fq(s) versus s for each value of q. For a stationary time series h(q=2) is identical with the well-defined Hurst exponent (Feder 1988). Again, for positive q, h(q) describes the scaling behavior of the segments with large fluctuations while for negative q value, h(q) describes the scaling behavior of the segments with small fluctuations. The generalized Hurst exponent h(q) of MF-DFA is related to the classical scaling exponent  $\tau$(q) by the relation
\begin{equation}
\tau(q)= qh(q)-1.
\end{equation}
A monofractal time series with long range correlation is characterized by linearly dependent exponent $\tau$(q) on q values with a single Hurst exponent. Multifractal signal has multiple Hurst exponents and $\tau$(q) depend nonlinearity on q. Furthermore, it is possible to characterize the multifractality of the time series by deriving the multifractal spectrum f($\alpha$), which is related to $\tau$(q) by a Legendre transform $\alpha$ = (d$\tau$)/(dq) and f($\alpha$) = q$\alpha$- $\tau$(q)(Feder 1988).

Where, $\alpha$ is the multifractal strength and f($\alpha$) specifies the dimension of subset series, that is characterized by $\alpha$. Using Eq. (13) we can write $\alpha$ and f($\alpha$) is related to h(q)
\begin{equation}
\alpha = h(q) + q(h'(q))
\end{equation}
\begin{equation}
F(\alpha)= q[\alpha-h(q)]-1.
\end{equation}

The multifractal spectrum generally quantifies the long-range correlation property of a time series. It is capable of providing information about the relative importance of various fractal exponents in the series, e.g. the width of the spectrum denotes range of exponents. It has been proposed by some workers that the width of the multifractal spectra is a measure of degree of multifractality (Ashkenazy et al. 2003). We can make a quantitative characterization of the spectra by least square fitting it to a quadratic function around the position of maximum $\alpha$o. The width of the spectrum can be obtained by extrapolating the fitted curve to zero. The width W is defined as
\begin{equation}
W = \alpha_{max}-\alpha_{min}.
\end{equation}

Multifractal dimension of subset series f($\alpha$) can be obtained from the relation (14) and (15). For a monofractal series, h(q) is independent of q. Hence, from the relation (14) and (15) it is evident that there will be a unique value of $\alpha$ and f($\alpha$), the value of $\alpha$ being the generalized Hurst exponent H and the value of f ($\alpha$) being 1. Hence, the width of the spectrum will be zero for a monofractal series. The larger the width, the richer the multifractality present in the dataset (Kantelhardt et al. 2002).

\section{Results and discussion}

\begin{figure*}[htb!]
\begin{center}
\includegraphics[height=4.5cm,width=16cm]{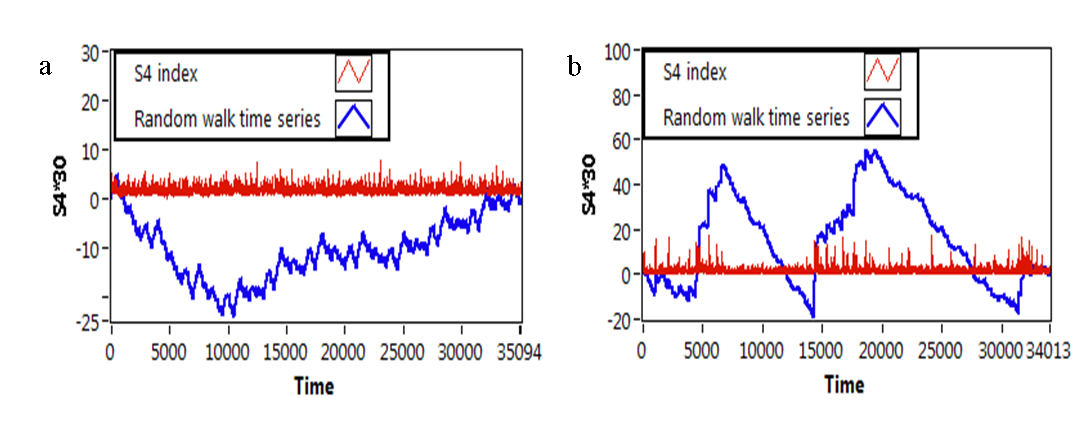}
\caption{Ionospheric scintillation $S_4$ index time series measured from the GSV4004B GISTM receiver at station Surat during $21^{st}$ March-2012 to $30^{th}$ March-2012 (red line) along with random walk time series (blue line), calculated by Eq. (4) under different time period (a) day time and (b) night time }
\label{Fig. 1}
\end{center}
\end{figure*}

By applying R/S, ACF and MF-DFA methods described in the Sec. 3, we analyze the self-affine and fractal properties of L-band amplitude scintillation index ($S_4$) time series, measured during the $21^{st}$ March-2012 to $30^{th}$ March-2012 under different time zones like day time (05:30-18:29) and night time (18:30-05:29).

\begin{figure*}[htb!]
\centering
\includegraphics[width=16cm,height=5cm]{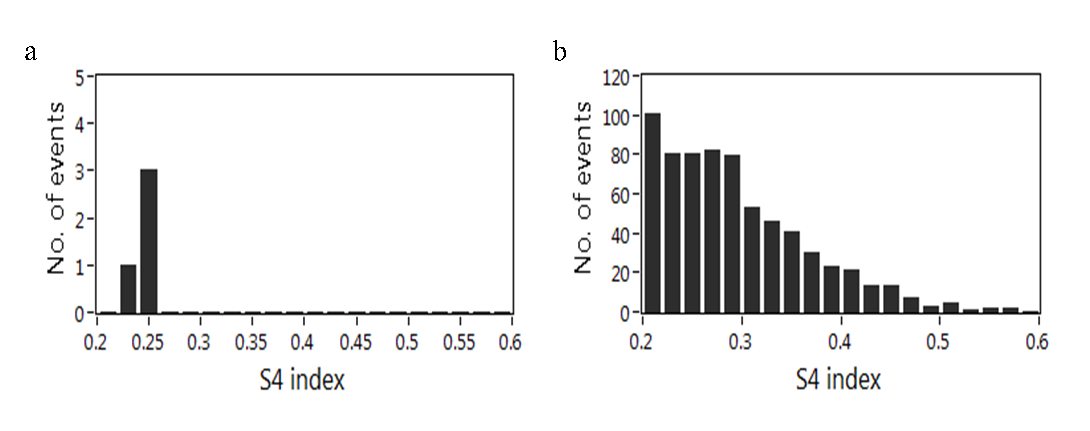}
\caption{Normal distribution of scintillation events in accordance with $S_4$ $\geq$ 0.2 for (a) day time and (b) night time}
\label{Fig. 2}
\end{figure*}

Fig. 1 shows day time and night time $S_4$ index time series along with random walk like integrated time series, derived from the Eq. (4). In the random walk time series, local daily trends in the data are eliminated by the subtraction of the local averages. By considering the detrended $S_4$ index time series, it is possible to obtain some information about the inhomogeneity of the ionosphere and the strength of GPS signal passing through it. Since we employ the conventional measurements to study the ionospheric scintillation, the corresponding time series consist of local fluctuation with both reasons: scintillation activity ($S_4$ $\geq$ 0.2), due to presence of plasma density irregularities or variations in TEC (Tanna et al. 2013) (large magnitude fluctuations), combined with the fluctuations due to internal dynamics of the system. As it is well reported that, low latitude ionospheric scintillations are mainly a post-sunset phenomena associated with the presence of F-region irregularities in the path of radio wave (Zou and Wang 2009; Zou 2011; Tanna et al. 2013 ), the distribution of events in accordance with $S_4$ $\displaystyle\geq$ 0.2 are thus plotted in the Fig. 2 for both time series. Fig. 2 shows the agreement of these phenomena as large no. of scintillation events observed during the night time. The random walk time series associated with the actual data series shown in the Fig. 1 reflects the self-similarity in the considered data sets as the similar trend observed after some data points (dos Santos Lima et al. 2012). In this case, it is important to confirm the type of self-affinity and fractal (multifractal) properties by estimating results over different time span s for the time series.

In order to characterize the type of self-affinity, the Hurst exponent has been calculated by two different methods: R/S and ACF. After calculating the standard deviation S(i) and range R(i) of both time series as discussed in section 3.1, we have calculated  R/S over time span s. Fig. 3 shows log-log plots of R/S of $S_4$ index fluctuations as a function of time span s, which estimate the values of Hurst exponent H = 0.7531 and 0.7242 for the day time and night time respectively. Moreover we apply the ACF method to the considered time series at time span s and plotted the natural logarithmic plots of ACF versus time span s in Fig. 4. According to Fig. 4, the values of Hurst exponents obtained by the Eq. (7) are 0.85 and 0.83 for the day time and night time $S_4$ index fluctuation respectively. The algebraic decay in the ACF and estimated values of Hurst exponents from the R/S and ACF methods reflect the long range correlation in the persistence data sets (Rangarajan and Ding 2000; Nurujjaman and Iyengar 2007).

\begin{figure*}[htb!]
\centering
\includegraphics[width=16cm,height=4.5cm]{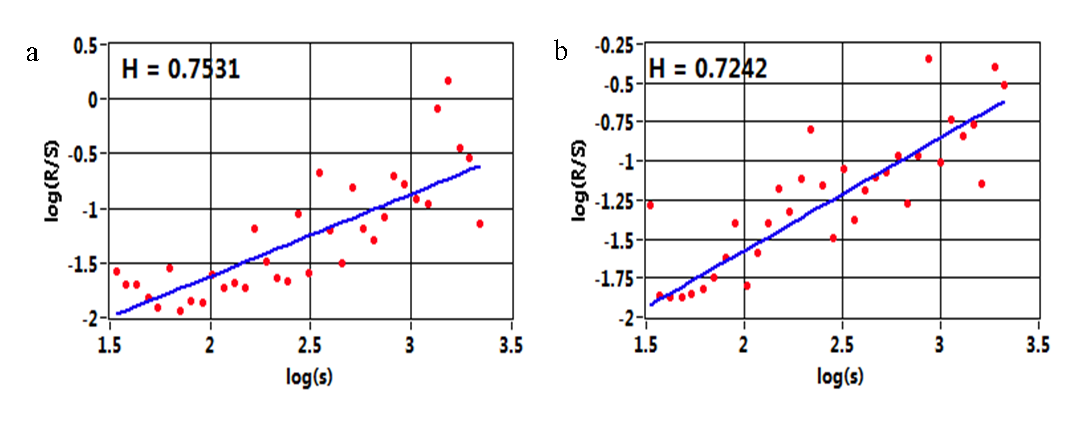}
\caption{Double logarithmic plots of (R/S) as a function of time lag for (a) day time (b) night time $S_{4}$ index time series}
\label{Fig. 3}
\end{figure*}

\begin{figure*}[htb!]
\centering
\includegraphics[width=16cm,height=4.5cm]{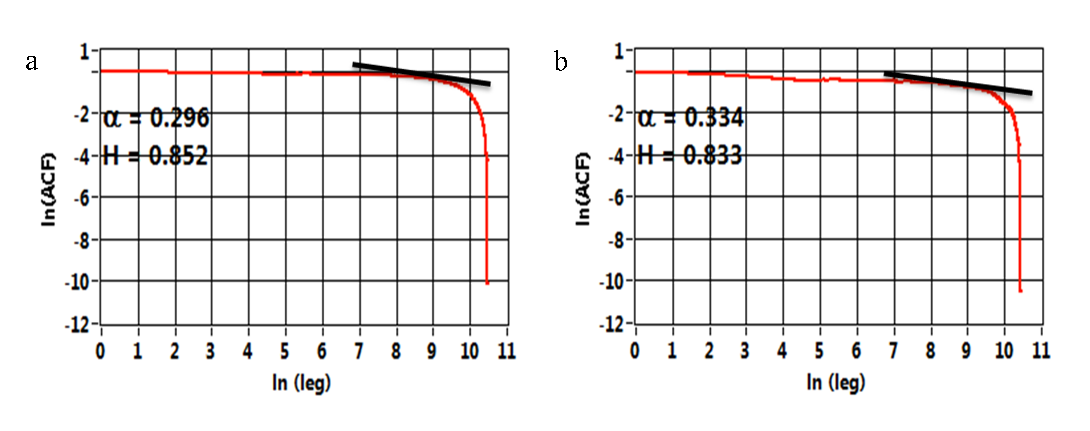}
\caption{ln-ln plots of ACF vs. time lag for (a) day time (b) night time $S_4$ index time series}
\label{Fig. 4}
\end{figure*}

\begin{figure}[htb!]
\centering
\includegraphics [width=7cm,height=10cm]{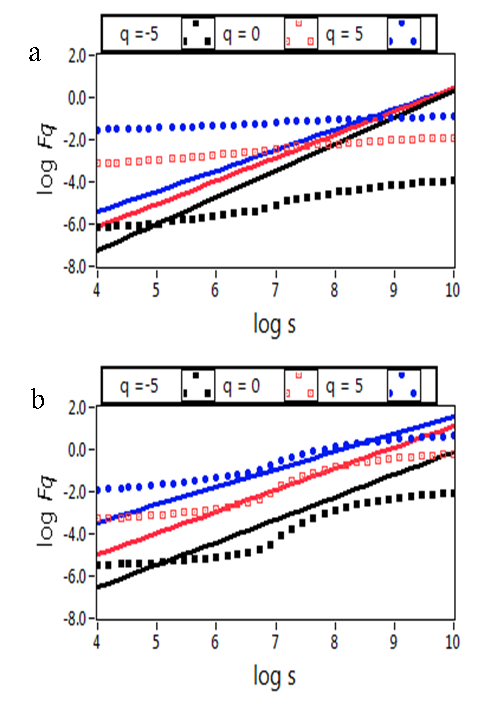}
\caption{The q-order fluctuation Fq and corresponding regression line for (a) day time and (b) night time}
\label{Fig. 5}
\end{figure}

In order to operate the considered time series by the MF-DFA method, we have first segmented integrated random walk time series into a number of bins (Ns) with different scale size. Then the the fluctuation function Fq(s) for q $\in$ [-10, 10] for all values of s from the Eq. (10) have been calculated. To investigate the behaviour of fluctuation function, the double logarithmic plots of Fq(s), for both time series as a function of s have been plotted, for three different q values, in Fig. 5. The difference between the $q^{th}$ order fluctuation for positive and negative q's at the smallest segment sizes compared to the large segment sizes for the both time series reveal the multiscalling properties of considered time series (Ihlen 2012). The small segments are able to distinguish between the local periods with large and small fluctuations (i.e., positive and negative q's, respectively) because smaller segments are embedded within these periods. However, the large segments cross several local periods with both small and large fluctuations and will therefore average out their differences in magnitude. The slope of the regression line of $q^{th}$ order fluctuation function and time span gives the well-known generalized Hurts exponent h(q). For monofractal data sets, the fluctuation function has a constant slope at any q values (Kantelhardt et al. 2002). While, in our case the slopes differ on q values and so infer that the L-band ionospheric scintillation signals has a multifractal temporal structure (Movahed et al. 2011).

\begin{figure}[htb]
\centering
\includegraphics[width=6.5cm,height=13cm]{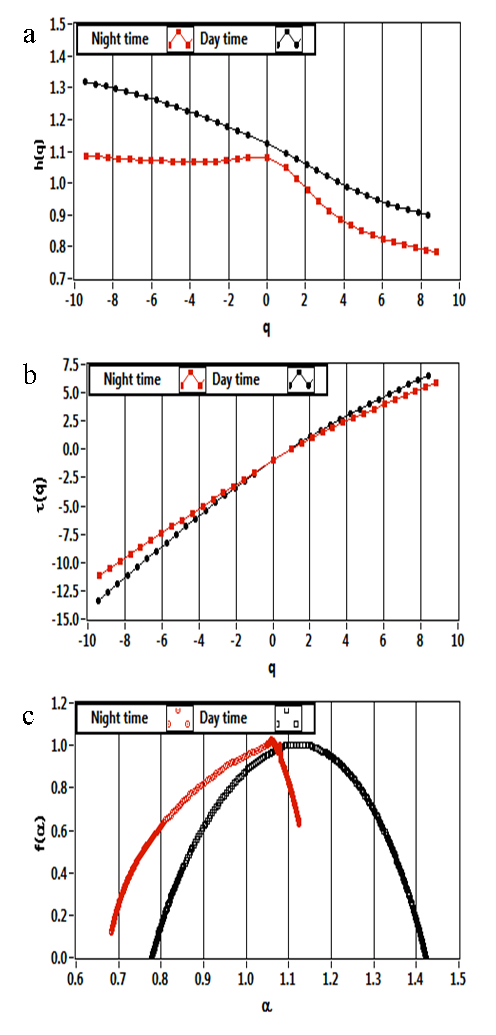}
\caption{(a) Generalized Hurst exponent h(q) (b) multifractal scaling exponent $\tau$(q) (c) Multifractal spectrum f($\alpha$) for the same selected day time (black line)  and night time (red line) $S_4$ index time series}
\label{Fig. 6}
\end{figure}

Besides, from the calculations of $q^{th}$ order fluctuation functions and the generalized Hurts exponent for the both: day time and night time data sets, the structure of low latitude ionosphere reveal the multi-scaling and multifractal properties. Thus, in order to verify the multifractal behavior in the dynamics of ionosphere, $q^{th}$ order generalized Hurst exponent h(q), classical multifractal scaling exponent (Reni exponent) $\tau$(q) and multifractal spectrum  f($\displaystyle \alpha$) have been calculated through Eq. (14), (15) and (16) respectively. Since Fig. (1-a) and (2-a) show the scintillation free day time ionosphere (i.e. normal ionosphere), the multifractal analysis of the day time data sets have been compared with the results obtained for the highly scintillated night time data sets. Fig. 6 shows the plots of h(q), $\displaystyle \tau$(q), and f($\displaystyle \alpha$) for the night time and day time $S_4$ index time series.

From the Fig. 6, for the day time $S_4$ index time series, the h(q) $\displaystyle\sim$ q relation is observed by the typical multifractal form, monotonically decreasing with the increase of q. This relation on h(q) with q leads to a nonlinear  $\displaystyle \tau$(q) dependence on q (Movahed et al. 2011). The nonlinearity indicates that there is non-linear interaction between the different scale events and multifractal nature of the system (Berge et al. 1988; Biskamp 1993). Consequently, the resulting multifractal spectra f($\displaystyle \alpha$) is a wide Gaussian like curve.

The absolute value of the width of the multifractal spectrum W calculated by the Eq. (16) as well as the shape of the multifractal spectrum is related to the temporal variation of the generalized Hurst exponent h(q) (Ihlen 2012). In this case, the multifractal spectrum width of the day time $S_4$ index time series is found to be 0.6533.

On the other hand, when the highly scintillated night time $S_4$ index time series is considered, the results are different. From the Fig.6 it is evident that the behavior of h(q) for q values and shape of the multifractal spectrum for the day time and night time are not similar. For the time series of day time, multifractal spectrum found to be symmetric (i.e. in Gaussian shape) whereas, during the night time, the spectrum was found with right-side truncation and thereby decreases the broadness of the spectrum. However, estimated values of the width of spectra are accordingly 0.6533 and 0.4419 (W $>$ 0) of the day time and night time data sets emphasize the presence of multifractality/complexity in considered system (Movahed et al. 2011; dos Santos Lima et al. 2012). The change in the shape of spectra for highly scintillated data may provide vision into the variations in the ionospheric behavior resulting from this scintillation analysis. Recently, Ihlen (2012) have reported that, the symmetric spectrum is originated from the levelling of the $q^{th}$ order generalized Hurst exponent for both positive and negative q's values. The levelling of q-order Hurst exponent reflects the q-order fluctuation is insensitive to the magnitude of local fluctuation. When the multifractal structure is sensitive to the small-scale fluctuation with large magnitudes, the spectrum will be found with right truncation; whereas, the multifractal spectrum will be found with left-side truncation when the time series has a multifractal structure that is sensitive to the local fluctuations with small magnitudes. In the present work, the right-side truncated spectrum of night time reflects the existence of small-scale intermittency in the post-sunset turbulent ionosphere. In the presence of enhanced eastward electric fields and meridional neutral winds during post-sunset time, the ionosphere becomes destabilized and causes the formation of plasma bubbles that penetrate into the topside ionosphere (Tanna et al., 2013) which, generate small-scale irregularities. This small-scale irregularities lead to intermittency in turbulent ionosphere. Intermittency implies a tendency of a physical entity to concentrate into small-scale features of a large magnitude of fluctuations surrounded by extended areas of less intense fluctuations (Monin and Yaglom 1975; Biskamp 1993; Frisch 1995). Thus, it may be noted here that, the width and shape of the multifractal spectrum is able to classify a small and large magnitude (intermittency) fluctuations in the considered time series, originated from the inhomogeneous ionosphere.

Compiling the h(q) and  $\displaystyle \tau$(q) behaviour and the obtained values of W, the results correspond to fingerprints of multifractal behaviour in the $S_4$ index time series, and complexity in the low latitude ionosphere.

\subsection{Comparison of the multifractal nature of the original, shuffled and surrogate $S_4$ index data}

To study the complexity of normal and turbulent ionosphere further, we performed two tests with surrogate and shuffled time series. In general, two different types of multifractality in certain data sets can be distinguished:

 \begin{enumerate}
      \item multifractality due to a broad probability distribution function (PDF) and
      \item multifractality due to different long-range correlation for small and large fluctuations.
 \end{enumerate}

\begin{figure*}[htb]
\centering
\includegraphics[width=16cm,height=9cm]{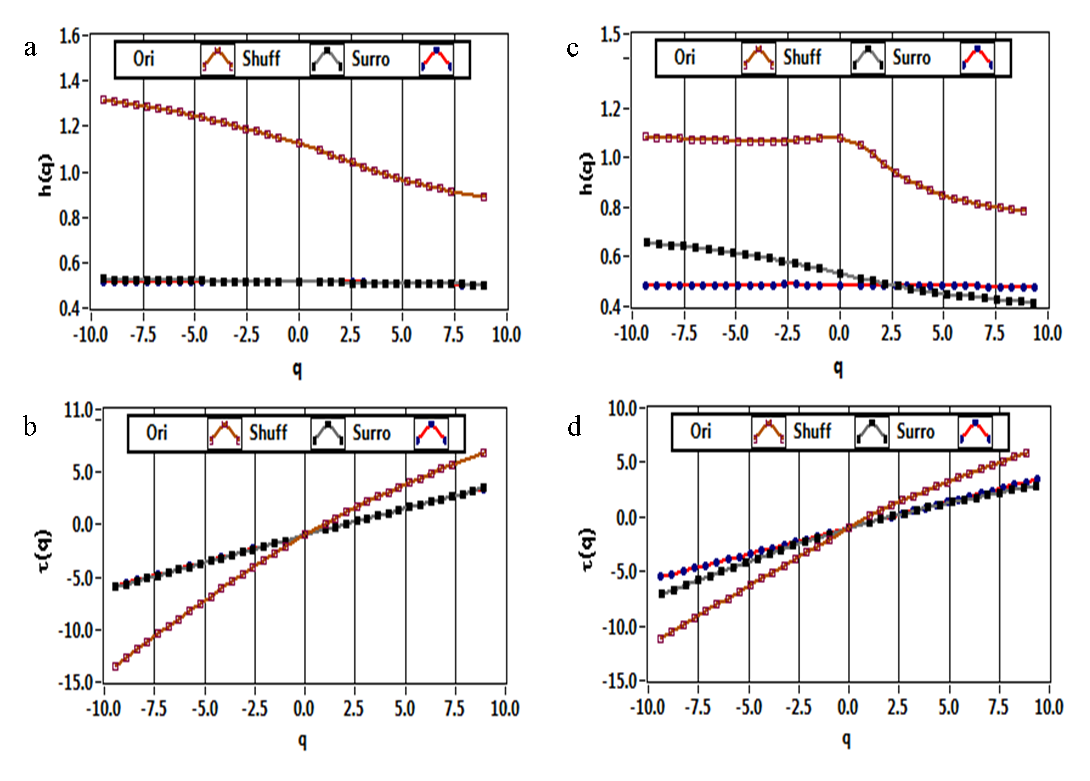}
\caption{(a) Generalized Hurst exponent h(q) (b) multifractal scaling exponent $\tau$(q) for the day time $S_4$ index time series of original (brown line), shuffled (black line)  and surrogate (red line), Similar plots of night time $S_4$ index time series of original, shuffled and surrogate in (c), (d)}
\label{Fig. 7}
\end{figure*}

\begin{figure*}[htb]
\centering
\includegraphics[width=16cm,height=5cm]{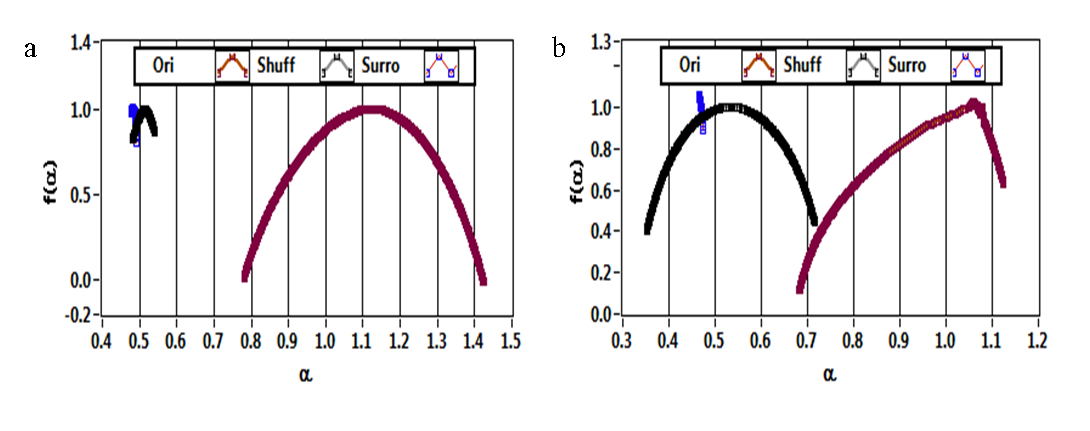}
\caption{Multifractal spectra f($\alpha$) for the same selected (a) day time  and (b) night time index time series of the original (pink line), shuffled (blue line) and surrogate (green line)}
\label{Fig. 8}
\end{figure*}

In first case the multifractality cannot be removed by shuffling time series, while in the second case data may have a PDF with finite moments and the corresponding shuffled series will reveal non-multifractal characteristic, as all long-range correlations are destroyed by the shuffling method. If both kinds of multifractality are present in the data sets, the shuffled series will show weaker multifractality than the original series (Kantelhard et al. 2002). The easiest way to identify the type of multifractality is by analysing the corresponding shuffled and surrogate data (Movahed et al. 2006).

The shuffling of time series destroys the long-range correlation. Hence, if the multifractality is only due to long-range correlation, the shuffled series will exhibit non-multifractal properties. The multifractal behaviour due to the broad PDF signals is not affected by the shuffling method. If the multifractality in the time series is due to wide PDF, h(q) obtained from the surrogate time series will be independent on q.

To identify the multifractal behaviour, we compare the fluctuation function Fq(s), for the original time series with the results of the corresponding shuffled, $F_{shuf}$(s)  and surrogate $F_{sur}$(s). The differences between these two fluctuation functions with of the original series indicate the presence of long-range correlation or the broadness of PDF in the original series. These differences can be obtained by the ratio $F_q$(s)/ $F_{shuf}$(s) and $F_q$(s)/ $F_{sur}$(s) as a function of time span s.  The scaling nature of these ratios is
\begin{equation}
F_{q}(s)/F_{q}^{shuf}(s)\sim s^{h(q)-h_{shuf}(q)}= s^{h_{corr}(q)}
\end{equation}
\begin{equation}
F_{q}(s)/F_{q}^{sur}(s)\sim s^{h(q)-h_{sur}(q)}= s^{h_{PDF}(q)}.
\end{equation}

If the source of multifractality is the broad PDF, one should find h(q) =  $h_{shuf}$(q) and   $h_{corr}$ = 0. Whereas, deviations from $h_{corr}$ = 0 indicate the presence of correlations, and q dependence of $h_{corr}$(q) indicates that multifractality is due to long-range correlation (Movahed et al. 2006).

\begin{table}[htb!]
\addtolength{\tabcolsep}{-3.5pt}
\caption{The values of Hurst exponent at(q=2) and width of the spectra for the original, shuffled and surrogate night-time and day-time $S_4$ index time series\label{table 1}}
\begin{tabular*}{8.4cm}{ccccc}
\tableline
\hline
&\multicolumn{2}{c}{\textbf{\underline{H(q=2)}}}&\multicolumn{2}{c}{\textbf{\underline{W}}}\\
%\cline{2 - 5}
Data & Day-time & Night-time & Day-time & Night-time\\
\tableline
Original &1.005 &0.965 &0.643  &0.442 \\
Shuffled &0.511 &0.477 &0.057 &0.363 \\
Surrogate &0.512 &0.486  &0.036  &0.022\\
\tableline
\hline
\end{tabular*}
%% Any table notes must follow the \end{tabular} command.
\end{table}

Fig. 7 shows the relation of h(q) and $\tau$(q) on q for the original, shuffled and surrogate day time and night time $S_{4}$ index time series. From these plots, it is observed that the multifractal nature of day time $S_{4}$ index time series is due to long-range correlation as monofractal behavior observed for both shuffled and surrogate data sets. Moreover, the multifractal spectra f($\alpha$), shown in Fig. 8 also show the similar reason for the multifractality in the day time ionospheric data, as the width of the shuffled and surrogate time series spectra are very smaller than that of the original time series. The values of the generalized Hurst exponent h(q=2) and width of the spectra for the original, shuffled and surrogate series are reported in Table 1.

In the case of night time, the observed behavior of h(q) on q for the shuffled time series reflects the multifractality nature of $S_{4}$ index time series is due to both broadness of PDF and long-range correlation. From the Table 1, it is found that the absolute value of $h_{corr}$(q) is greater than $h_{PDF}$(q), hence multifractality due to broad PDF is weaker than that the multifractality due to the correlation (Movahed et al. 2006). On the other hand, the width of the singularity spectrum, f($\alpha$) for original, shuffled, and surrogate time series, are approximately 0.442, 0.363 and 0.022 respectively. These values also show that the multifractality due to long-range correlation is dominant. Furthermore, calculated values of Hurst exponents through the R/S and ACF procedures for both considered time series, also suggest that the multifractal nature in the ionospheric fluctuations is due to long-range correlations.

\section{Conclusion}
This is the first ever attempt to analyze the fractality of the ionospheric scintillation time series obtained at an Indian low latitude station. In this work, we investigate the fractal properties of day-time and night-time $S_4$ index time series by using three different methods. The computed values of Hurst exponents by the R/S and ACF methods are $\in$ [0.5, 1] show that the existence of long range correlation in persistence ionospheric scintillation data. The MF-DFA method allows us to determine the non-linearity and complexity of ionospheric dynamics.The q dependence of generalized Hurst h(q), classical scaling $\tau$(q) exponents and the broadness of the spectra indicate that the $S_4$ index time series has multifractal behavior. By comparing the results of the original time series with that of the shuffled and surrogate ones, we found that the multifractality due to correlation makes a greater contribution than the broad PDF. According to the difference between the generalized Hurst exponent and the shape of the spectra of the day time and night time data sets, we conclude that MF-DFA is able to classify a wide range of different scale invariant structure of inhomogeneous and intermittent ionosphere. Furthermore, these results indicate that the quiet ionospheric scintillation data is even more complex than that of the highly scintillated ones, posing challenges to ongoing efforts to develop realistic models of mitigation of scintillation measurements and other processes under inhomogeneous low latitude ionosphere.

\acknowledgments
The authors acknowledge the support through TEQIP grant, received from the World Bank for the procurement of GPS system at Surat. The first author also acknowledges the fellowship support extended to her by the S. V. National Institute of Technology.

\end{document}